\documentclass[10pt,letterpaper]{article}

\newcommand{\unit}[1]{\ensuremath{\;\mathrm{#1}}}

\usepackage{opex3}
\usepackage{graphicx}

\begin{document}

\title{Entangled quantum key distribution over two free-space optical links}

\author{C. Erven,$^1$ C. Couteau,$^1$ R. Laflamme,$^{1,2}$ and G. Weihs$^1$}

\address{$^1$ Institute for Quantum Computing and Department of Physics and Astronomy, University of Waterloo, 200 University Avenue West, Waterloo, ON, N2L 3G1, Canada}
\address{$^2$ Perimeter Institute, 31 Caroline Street North, Waterloo, ON, N2L 2Y5, Canada}

\email{cerven@iqc.ca} 



\begin{abstract}
We report on the first real-time implementation of a quantum key distribution (QKD) system using entangled photon pairs that are sent over two free-space optical telescope links. The entangled photon pairs are produced with a type-II spontaneous parametric down-conversion source placed in a central, potentially untrusted, location. The two free-space links cover a distance of 435\unit{m} and 1,325\unit{m} respectively, producing a total separation of 1,575\unit{m}. The system relies on passive polarization analysis units, GPS timing receivers for synchronization, and custom written software to perform the complete QKD protocol including error correction and privacy amplification. Over 6.5 hours during the night, we observed an average raw key generation rate of 565\unit{bits/s}, an average quantum bit error rate (QBER) of 4.92\%, and an average secure key generation rate of 85\unit{bits/s}.
\end{abstract}



\section{Introduction}

Quantum key distribution (QKD) has become one of the first mature applications to develop out of the new field of quantum information processing. From the initial ideas of uncloneable quantum money by Wiesner \cite{Wie83} in the 1970's, to the first concrete QKD protocol (BB84) discovered by Bennett and Brassard \cite{BB84} in 1984; QKD has rapidly become a very practical application of quantum information science. There are now a number of different QKD protocols which have been demonstrated using both optical fibers and free-space optical links as their quantum channel. Some of the more recent free-space experiments include the distribution of entanglement through intra-city free-space links by Resch \emph{et al.} \cite{RLBBFKPSTUWWWZ05} and Peng \emph{et al.} \cite{PYBZJFYYYZLTP05}, the distribution of entanglement over 144\unit{km} and subsequent generation of a secure key by Ursin \emph{et al.} \cite{UTSWSLBJPTOFMRSBWZ07}, and the complete implementation of a free-space QKD system which included all key extraction routines by Marcikic \emph{et al.} \cite{MLK06}. For a comprehensive overview of both the theory and different experimental implementations of QKD please refer to the recent review article by Scarani \emph{et al.} \cite{SBCDLP08}.

While fiber implementations have produced some of the fastest systems to date, until reliable quantum repeaters are realized, fiber implementations seem to be limited to $<$ 200km. This has prompted increased attention on free-space implementations. Indeed, a number of studies have been performed to evaluate the possibility of performing quantum key distribution with an orbiting satellite such as the International Space Station \cite{RTGK02,AJPLZ03,PFMMCPAJUSBRLBWZ07}. Therefore, experience with free-space quantum key distribution in a variety of setups and experimental conditions is very valuable for future long distance experiments.

\section{Security assumptions}
Although the unconditional security of many QKD protocols has been shown \cite{SBCDLP08}, practical implementations are always different from the ideal theory and the possible presence of side channels require that great care is taken when claiming that one has implemented an unconditionally secure quantum key distribution system. Certain assumptions, which are required for the security proofs, are not always met in practice. To that end, we state fully here the assumptions going into the claim of security for our QKD system.

First, to prevent a man-in-the-middle attack, all classical communications between Alice and Bob must be authenticated using a short amount of initial secret key for the first few messages and by key generated by the system afterwards. The key generation process is still efficient because the number of bits needed for authentication is logarithmic in the size of the key \cite{ABBDDGGGLLLPPPPRRRRSSWZ07}. We have not implemented authentication in our system yet, but it is under development.

Second, as will be shown in the discussion of our results, the detection efficiencies for each of Alice's and Bob's detectors are not equal. Attacks are known that can exploit a detector inefficiency mismatch \cite{QFLM07} and perhaps leave our system vulnerable to an eavesdropper. The solution is to carefully equalize the efficiencies of all the detectors without exposing further security loopholes. We are currently investigating the methods to properly equalize the detector efficiencies.

Third, spatial filtering was not used in the polarization detector boxes described below. It is known that if different detectors see slightly different spatial modes, an eavesdropper could control their relative efficiencies by varying the spatial mode of the input signal \cite{ML08}. However, a significant effort was expended to make the relevant dimensions in the detector box symmetric and identical in order to avoid precisely this kind of attack. Strictly speaking this attack might still be possible, but is assumed to be minimal for our system.

Fourth, double clicks, that is when two detectors register a photon at the same time, need to be kept track of and should be assigned a random bit value. For entanglement based QKD it is unclear whether double pair emissions from the source lead to the same drastic security loophole experienced by weak coherent pulse QKD; namely, the photon number splitting attack. Nevertheless, it is important to keep track of these events and assign a random measurement result when a double click is observed. In the current system, we do not explicitly deal with double clicks, but the system will choose whichever event it records first. This is similar to a random choice within the detector time jitter.

Fifth, we make certain assumptions about our detectors, namely that each of the four channels in our detectors are independent and that the dead-time of the detectors is a negligible security concern. The assumption that each of the four channels in our detectors are independent means that the probability of detecting a photon on any channel is independent of whether a photon was previously detected on any other channel. While this is a natural assumption to make, it is not always found in practice and depends on the electrical design and physics of the detectors. Additionally, each channel has a certain dead-time after a detection event, in our case 50\unit{ns}, where the channel cannot detect another photon until it has been reset. To ensure security, it is important to reject multiple detection events that fall within the dead-time of the detectors. While these loopholes were not pointed out to us until after the experiment was performed, we assume that their influence on the security of our system is negligible since the rate of detected photons is low enough that any lingering effects of one channel firing should have dissipated before the next channel detection event and the probability of having two detection events in the dead-time window was extremely small. A data rejection algorithm for multiple events being registered within the dead-time window will be added to future versions of the system.

Finally, we assume that the security proof by Ma \emph{et al.} \cite{MFL07} applies to our system, since it is the closest proof to our experimental implementation. However, it does not precisely encompass our implementation on its own since it assumes the validity of the squashing model of detection and that active basis switching is performed. However, we can make it apply to our system since the validity of the squashing model for the active basis switching detection scheme has recently been shown \cite{BML08,TT08,KAYI08}. Additionally, it has also recently been shown that the requirement for active basis switching can be relaxed to include the passive scheme, which is used in this experiment  \cite{Lut08}. We thus inherit the same two assumptions used in their proof when claiming security for our system. The assumptions are that we are operating in the long key limit, and that the bit and phase error rates can be assumed to be equal. Preliminary results \cite{HHHTT07} suggest that the overhead is indeed dramatic for finite key statistics and many more bits are needed than the formulas in many security proofs suggest. Nevertheless, secure key generation with the rates observed with our system should still be within the realm of possibility. The proof also makes the simplification that the bit and phase error rates are equal which needs to be carefully examined for our system.

\section{Experimental implementation}
We have built the first real-time implementation of a two free-space link entanglement based quantum key distribution system using the BBM92 protocol invented by Bennett \emph{et al.} \cite{BBM92}. Our system is comprised of a compact spontaneous parametric down-conversion (SPDC) source, two free-space telescope links, two compact passive polarization analysis modules, avalanche photodiode (APD) single photon detectors, time-stampers, GPS time receivers, two laptop computers, and custom written software \nocite{WE07} \nocite{Erv07} \cite{Erv08}.

\begin{figure}[htbp]
    \centering
    \includegraphics[width=13cm]{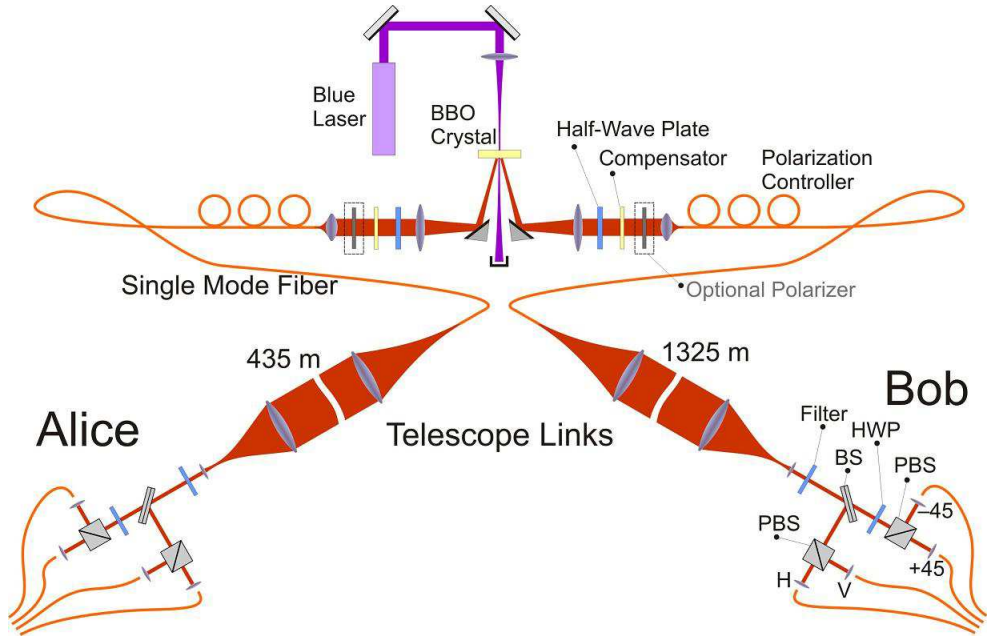}
    \caption{Experimental schematic of the entangled photon source, free-space link optics, and the passive polarization detection optics. Polarization entangled photons are generated via type-II spontaneous parametric down-conversion in a $\beta$-BBO nonlinear optical crystal pumped by a blue diode laser. Walk off effects are mitigated with a half waveplate and compensator crystal in each arm. Optional polarizers allow the measurement of the local source visibilities. The entangled photons are coupled into singlemode fibers and transported to sender telescopes where they are then sent over a free-space link and collected with receiver telescopes. The photon polarizations are then measured in a passive polarization detector box which uses a 50/50 beamsplitter to perform the basis choice and the proper combination of half waveplates and polarizing beamsplitters to perform the polarization measurement in the correct basis. \label{fig.EntangledSourceSchematic}}
\end{figure}

Entangled photon pairs are generated via a compact type-II spontaneous parametric down-conversion (SPDC) source \cite{KMWZSS95}, which was built on an optical breadboard measuring 61\unit{cm} by 46\unit{cm}. A schematic of the source is shown in Fig. \ref{fig.EntangledSourceSchematic}. The source is pumped by a 50\unit{mW}, 407.5\unit{nm} cw violet diode laser from Blue Sky Research that is focused to an approximate radius of 25$\unit{\mu m}$ in a 1\unit{mm} thick $\beta$-BBO crystal. The down-converted photon pairs at a degenerate wavelength of 815\unit{nm} are split off via two small prism mirrors. An achromatic doublet lens (f = 150\unit{mm}) collimates the down-converted photons and a half waveplate oriented at $45^{\circ}$ plus a 0.5\unit{mm} $\beta$-BBO crystal in each arm compensate for longitudinal and transverse walk-off effects. The angle of one of the compensator crystals is also used to set the relative phase between horizontal and vertical polarizations in order to produce the singlet Bell state. After compensation, the photons are coupled into short singlemode optical fibers using aspheric lenses (f = 11\unit{mm}), which can then be coupled either to long singlemode fibers which will carry the photons to the sending telescopes or to local detectors. The fibers pass through manual polarization controllers which are used to undo the polarization rotation induced by the singlemode fibers.

For local alignment, the short singlemode fiber is connected to a singlemode fiber - air - multimode fiber bridge which contains a narrowband spectral filter, centred at 815\unit{nm} with a 10\unit{nm} bandwidth (FWHM), in order to get rid of any residual laser light and background light before connecting the fibers to APD single photon detectors (PerkinElmer). These are the same filters which are used in the polarization analysis modules. Connecting the fibers to the detectors with this method and inserting the optional polarizers (see Fig. \ref{fig.EntangledSourceSchematic}) mounted on flip mounts allows us to measure the local quality of the entangled photon source. Typically, we measure a pair rate of 12,000\unit{s^{-1}} and total single photon count rates on each side of 100,000\unit{s^{-1}}. The local entanglement quality is ascertained by measuring the visibility of the source in the rectilinear (H/V) basis and the diagonal ($+45^{\circ}$/$-45^{\circ}$) basis. For the experimental run detailed in this article we measured visibilities of 99.6\% and 91\% respectively shortly before the start of the experiment. This corresponds to a local QBER of 2.35\%. The limited visibility in the diagonal basis is likely due to the broad spectral filtering (10\unit{nm}) and uncompensated transverse walk-off in the $\beta$-BBO crystal which is aggravated by the narrow pump beam spot.

\begin{figure}[htbp]
    \centering
    \includegraphics[width=13cm]{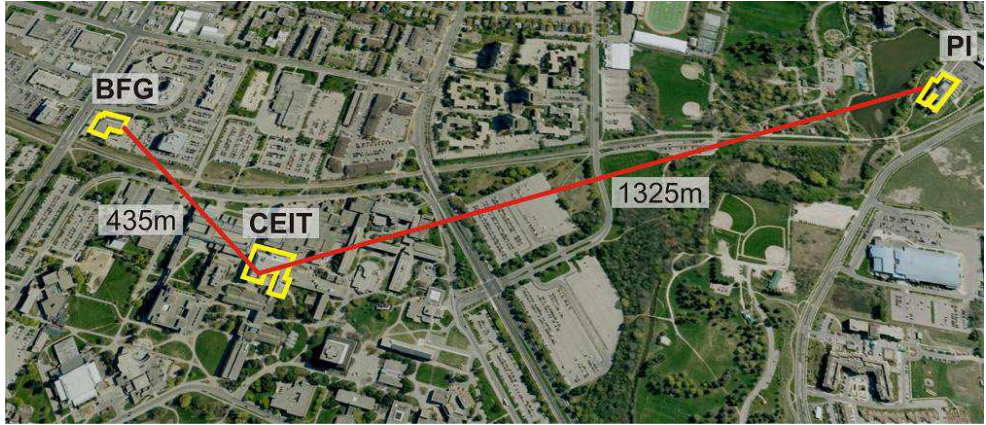}
    \caption{Map of the QKD setup showing the University of Waterlo and Perimeter Institute campuses with the source located in the CEIT building, Alice located 435\unit{m} away in an office at the BFG building, and Bob located 1,325\unit{m} away in an office at PI. Courtesy of Google Earth and Tele Atlas. Map Data \copyright \hspace{1 pt} Tele Atlas 2008\label{fig.Map}}
\end{figure}

For an experimental run, the short singlemode fibers are connected to longer 30\unit{m} singlemode fibers which transport the photons to two telescopes situated in telescope enclosures on top of the CEIT building shown in Fig. \ref{fig.Map}. The sender telescopes consist of a fiber adapter, which holds the end of the singlemode fiber along the optic axis and allows the light to expand and be collimated by an achromatic doublet lens (f = 250\unit{mm}, d = 75\unit{mm}) into an approximately 50\unit{mm} beam. The fiber adapter is mounted onto a translation stage driven by a high-resolution stepper motor which can adjust the focusing of the telescope. The same motors are used in the mount which holds the sender telescope to adjust its azimuthal and elevation angles. All the motors can be controlled remotely from an operator at Alice's or Bob's location in order to align the sender telescope with the receiver system.

The source location is a potentially untrusted location with the receivers situated at two distant locations with no direct line of sight between them (see Fig. \ref{fig.Map}). Alice's receiver sits in an office at BFG a free-space distance of 435\unit{m} away from the source. Bob, on the other hand, sits in an office at PI 1,325\unit{m} away from the source. This produces a total separation of 1,575\unit{m} between Alice and Bob.

The receiver system consists of a receiver telescope with a passive polarization detector box, that is mounted onto a homemade precision tip/tilt stage for fine adjustment of the receiver's pointing. The receiver telescope consists of an achromatic doublet lens (f = 200\unit{mm}, d = 75\unit{mm}) and a second small lens (f = 10\unit{mm}, d = 5\unit{mm}) which collimates the photons down into a beam approximately 3\unit{mm} in diameter. The beam then passes through a narrowband spectral filter (described above) to remove as much background light as possible and into the passive polarization analysis box shown in the bottom portion of Fig. \ref{fig.EntangledSourceSchematic}. A 50/50 nonpolarizing beamsplitter performs the basis choice by randomly reflecting or transmitting an incoming photon. Measurement of the photons in the diagonal basis is performed by a half waveplate and a polarizing beamsplitter in the transmitted arm; while measurement of the photons in the rectilinear basis is performed in the reflected arm with only a polarizing beamsplitter. The photons are then collected into four multimode fibers with permanently mounted aspheric lenses (f = 11\unit{mm}).

Tests of the polarization detector boxes revealed the typical leakage of some horizontally polarized photons into the vertical channel at the polarizing beamsplitters which would lead to an increase in the QBER rate of up to 1.5\%. The average local QBER rate for our source combined with this error in the polarization analysis boxes yields a baseline average QBER of 3.85\%.

Finally, the photons delivered through the four multimode fibers are detected by a quad single photon counting module from PerkinElmer which has an approximate detection efficiency of 50\% at our wavelength. In order to make the system simpler and remove the need for a separate timing channel to identify pairs of entangled photons, each photon detection event is time-stamped by units developed by Dotfast Consulting which have a time resolution of 156.25\unit{ps}. Since the photon pairs are created at the same time in the $\beta$-BBO crystal, entangled photon pairs correspond to simultaneous detection events after path length differences are taken into account. Accepting only simultaneous detections reduces the background almost to zero. However, this requirement currently forces us to experiment at night, since the background detection rates experienced during the day both overload our photon detectors and make entangled photon identification with this method infeasible.

At each location, GPS timing units from Spectrum Instruments provide a highly accurate 10\unit{MHz} reference signal to the time-stamping units. A one pulse per second (1PPS) signal provides a means to continually re-synchronize the electronics at Alice's and Bob's locations automatically, allowing indefinite stable timing operation of the whole system. Detection data is then passed via a USB connection to a laptop computer at Alice's or Bob's location, which then performs the classical parts of the BBM92 protocol.

At the beginning of an experiment, Alice's and Bob's computer clocks are first synchronized to $<$100\unit{ms} using a NIST timing application \cite{NISTTime} in order to give them a relatively accurate common start time. A measurement program thread is responsible for continually processing detection events and sending data on to a coincidence thread. The coincidence thread then exchanges timing information for Alice's and Bob's detection events in order to identify entangled photon pairs. A coincidence histogram is calculated to determine the timing offset between Alice's and Bob's measurements and then coincident detection is performed using a coincidence window of 2\unit{ns} in order to identify entangled photon pairs. At this point, Alice and Bob now have raw key data corresponding to entangled photon detection events. Along with the timing information, Alice and Bob also exchange measurement basis information for each detection event. This allows the coincidence thread to sift the raw key data down to only those detection events where Alice and Bob measured in the same basis yielding the sifted key. All of the classical communication is performed over an ordinary classical internet connection.

Ideally, Alice and Bob would now share identical keys which they could use to encrypt data; however, due to imperfect state production, transmission, and polarization analysis, not to mention any intervention by an eavesdropper, we expect to see errors in the sifted key. Errors are removed by performing a modified cascade error correction algorithm \cite{BS94} on the sifted key. Cascade uses public discussion to compare the parities of randomly chosen blocks from the sifted key and then performs a binary search on any blocks where the parities differ in order to identify and correct the error. It uses a multi-pass strategy in order to correct all errors with a high probability. During error correction, each parity communicated essentially leaks one bit of information to any eavesdropper monitoring the classical communication channel. The number of bits revealed during error correction is noted so that it can be taken care of in the privacy amplification stage.

The last step is for Alice and Bob to perform privacy amplification to reduce the amount of information an eavesdropper might possibly know about the key to an exponentially small amount at the cost of reducing the size of their key somewhat. First, Alice and Bob must calculate the fraction of their raw, error free key which they will be able to keep after privacy amplification. The calculation for this comes from the proof of security which most closely matches our physical implementation \cite{MFL07}. It bounds an eavesdropper's information as a function of the QBER and estimates the necessary key reduction factor using Eq. \ref{eq.BitRate}
\begin{equation}\label{eq.BitRate}
N_{\rm{secure}} = N_{\rm{raw}}(1-h_{2}(\rm{QBER})) - N_{\rm{leakage}} - N_{\rm{safety}}
\end{equation}
where $N_{\rm{secure}}$ is the final number of secure bits which Alice and Bob will have after privacy amplification, $N_{\rm{raw}}$ is the number of bits after error correction, $h_{2}(x) = -x \log x - (1-x) \log (1-x)$ is the binary entropy function, $N_{\rm{leakage}}$ is the number of bits revealed during error correction, and $N_{\rm{safety}}$ is an additional safety parameter, which we set to 30 bits in our experiments. Using this reduction ensures that our system is secure both against symmetric individual attacks (QBER $<$ 14.6\%) and coherent attacks (QBER $<$ 11\%); generating no key if the QBER rises above 11\%. Note that it is possible to achieve secure key distribution with QBER's above 11\% but it requires the use of two-way classical post-processing which we do not perform in our system. Thus, the upper limit of secure key generation for our system is a QBER of 11\%.

Alice and Bob then perform privacy amplification by applying the 2-universal hash function \cite{CW79} shown in Eq. \ref{eq.PrivacyAmp}
\begin{equation}\label{eq.PrivacyAmp}
k_{\rm{secure}} = (m \cdot k_{\rm{corrected}} + n) \bmod p
\end{equation}
to the raw error corrected key and keeping the last $N_{\rm{secure}}$ number of bits from the end. Here $k_{\rm{secure}}$ is the final secure key, $k_{\rm{corrected}}$ is the error corrected key, $m$ and $n$ are large random numbers (smaller than $p$) generated by a random seed shared by Alice and Bob, and $p$ is a large prime number. Alice and Bob can verify that they have obtained the same key after these information reconciliation steps by checking the random hash values of their strings many times.

\section{Results}
The experiment detailed below was performed on April 28, 2008. At the beginning of the experiment, the two free-space links were initially aligned with a red laser diode (658\unit{nm}) coupled into a singlemode fiber, connected to the sending telescope, and sent over the free-space link. After the shorter 435\unit{m} BFG link, this produced a spot at the receiver approximately 30\unit{mm} in diameter that typically wandered less than 10\unit{mm} from its centre. Airy rings were clearly visible in the spot over this shorter distance. The spot produced after the 1325\unit{m} PI link was significantly worse, with a diameter of approximately 100\unit{mm} and a typical wander of 50 to 100\unit{mm} from its centre. Additionally, Airy rings were rarely visible in the spot indicating a significant amount of scintillation in the beam. A significant amount of the drastic degradation over the longer link can be attributed to the fact that the beam passes over an exhaust vent shortly after leaving the sender telescope.

Across the shorter BFG free-space link we received about 29,000\unit{photons/s} from the source, while we received about 10,000\unit{photons/s} from the source across the PI link. Taking into account the detection efficiency of 64\% for Alice's polarization detector box and 60\% for Bob's polarization detector box \cite{Erv07}, this yields a link transmission efficiency of approximately 45.3\% for Alice and 16.7\% for Bob. The average detection rates for each of Alice's and Bob's detectors are shown in Table \ref{tab.DetectorRates} including an estimate of the counts due to background light, dark counts, and photons received from the source.

\begin{table}[htbp]
  \centering
  \begin{tabular}{|c|c|c|c|c|c|c|c|c|}
      \hline
       & \multicolumn{8}{|c|}{Average Detection Rates (photons/s)} \\
      \hline
       & H & V & + & - & Total & Background & Dark Counts & Source \\
      \hline
      Alice & 5,203 & 7,755 & 6,741 & 6,194 & 25,893 & 14,693 & 1,200 & 10,000 \\
      \hline
      Bob & 7,845 & 12,296 & 11,219 & 10,980 & 42,340 & 12,140 & 1,200 & 29,000 \\
      \hline
  \end{tabular}
  \caption{Detection Rates for each of Alice's and Bob's detectors including an estimated breakdown of events due to background light, dark counts of the detectors, and photons received from the source. \label{tab.DetectorRates}}
\end{table}

\begin{figure}[htbp]
    \centering
    \includegraphics[width=13cm]{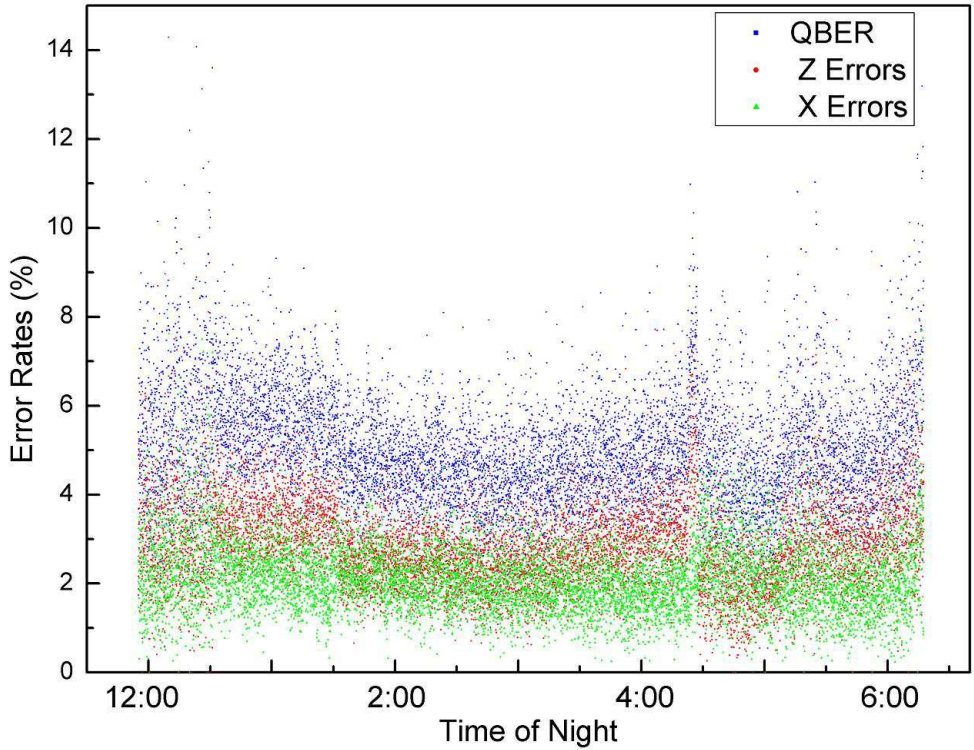}
    \caption{QBER over the course of the experiment. The total QBER is shown in blue, the Z errors are shown in red, and the X errors are shown in green. We observed an average total QBER of 4.92\% during the night. Data taken on April 28, 2008. \label{fig.QBER}}
\end{figure}

During the experiment we observed an average coincidence rate of 565\unit{s^{-1}} which varied wildly due to the beam fluctuation over the PI link. Fig. \ref{fig.QBER} shows the quantum bit error rate (QBER) observed over the course of the experiment from 11:55 pm until 6:15 am at which point the rising sun saturated our detectors and made correct coincidence detection impossible due to the high background. This caused the QBER to skyrocket and prevented further secure key generation. The contributions to the total QBER from both X and Z errors are also shown. The total average QBER during the experiment was observed to be 4.92\% of which 2.11\% and 2.81\% were X and Z errors respectively. The increase in the QBER from the baseline 3.85\% expected to the observed 4.92\% is due to residual uncompensated birefringence in the singlemode fiber used to transport the photons from the source to the sender telescopes and to accidental coincidences.

\begin{figure}[htbp]
    \centering
    \includegraphics[width=13cm]{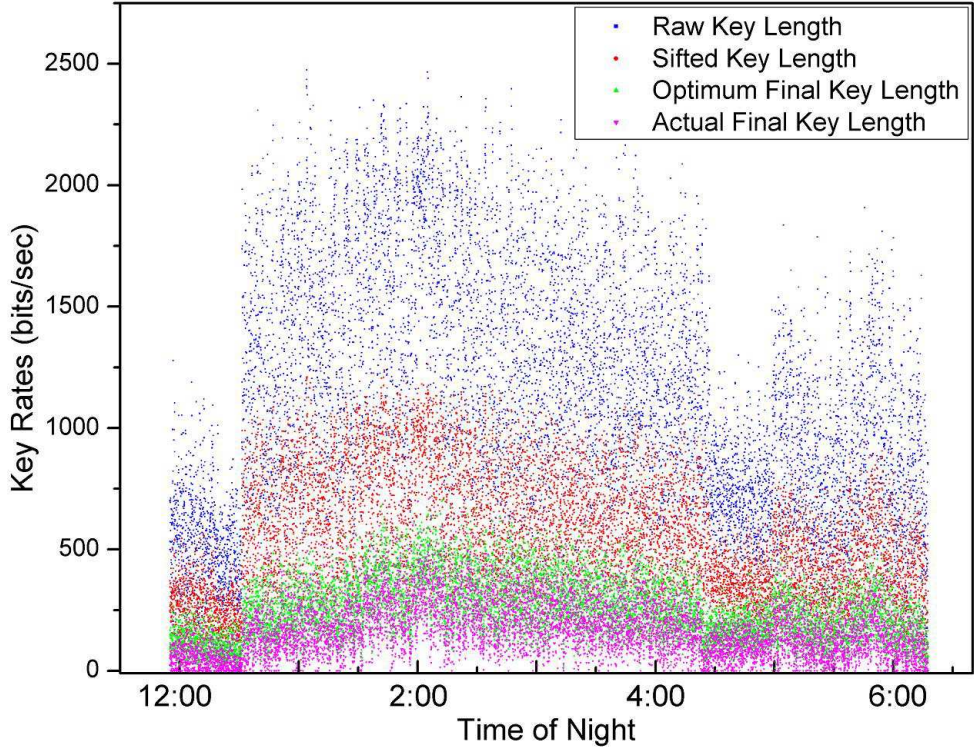}
    \caption{Key Rates over the course of the experiment. The raw key rate is shown in blue, the sifted key rate is shown in red, the maximum potential final key rate using error correction algorithms operating at the Shannon limit is shown in green, and the actual observed final key rate is shown in magenta. During the experiment we observed average rates of 565\unit{bits/s} for the raw key, 284\unit{bits/s} for the sifted key, 124\unit{bits/s} for the optimum final key, and 85\unit{bits/s} for the actual final key. Data taken on April 28, 2008. \label{fig.KeyRates}}
\end{figure}

Fig. \ref{fig.KeyRates} shows the key rates observed during the experiment with the raw key rate shown in blue, the sifted key rate in red, the theoretical maximum possible final key rate secure against coherent attacks (QBER $<$ 11\%) in the model from \cite{MFL07} with an error correction algorithm operating at the Shannon limit in green, and the actual observed final key rate shown in magenta. The jump in raw key rate around 12:30 am is due to changing the collection time from one second to two seconds for each data point which was necessary due to the low count rates in order to maintain a software coincidence lock. Further drops in key rates though were due to the system slowly becoming misaligned during the night.

We observed an average raw key rate of 565\unit{bits/s}, an average sifted key rate of 284\unit{bits/s}, an average optimal final key rate of 124\unit{bits/s}, and an average actual final key rate of 85\unit{bits/s}. As can be seen in Fig. \ref{fig.KeyRates}, the final key rate for our system was below the theoretical limit due to the fact that our classical post-processing does not operate at the Shannon limit. The experiment generated a total raw key of 10,806,880\unit{bits}, a total sifted key of 5,422,762\unit{bits}, a maximum possible optimal final key of 2,374,384\unit{bits}, and an actual final key of 1,612,239\unit{bits}. In other words, the experiment was able to generate over 200\unit{kB} of secure key during the night.

Table \ref{tab.CoinMatrix} shows the reconstructed coincidence matrix from Alice's and Bob's measurement data recorded during the experiment. The detection totals for Alice's and Bob's measurements of H, V, +, and - are also displayed and show an obvious variation in the detection efficiencies for each channel. As was discussed earlier, detector efficiency mismatches open a loophole, which an eavesdropper can exploit. Work to properly equalize the detector efficiencies is currently underway.

\begin{table}[htbp]
  \centering
  \begin{tabular}{|c|c|c|c|c|c|c|}
      \hline
       & & \multicolumn{4}{|c|}{Alice} & \\
      \hline
       & & H & V & + & - & Total\\
      \hline
       & H & 39,497 & 1,218,454 & 393,100 & 355,074 & 2,006,125 \\
      \hline
      Bob & V & 1,300,749 & 112,793 & 682,595 & 854,848 & 2,950,985 \\
      \hline
       & + & 680,032 & 878,628 & 51,217 & 1,262,143 & 2,872,020 \\
      \hline
       & - & 548,695 & 955,146 & 1,374,648 & 63,261 & 2,977,750 \\
      \hline
       & Total & 2,604,973 & 3,165,021 & 2,501,560 & 2,535,326 & \\
      \hline
  \end{tabular}
  \caption{Reconstructed coincidence matrix for Alice and Bob from the experiment. Also shown are the entangled photon detection totals for each of Alice's and Bob's four detectors. \label{tab.CoinMatrix}}
\end{table}

As an intermediate step, we performed a first estimate of the extra privacy amplification needed to take care of the unequal \emph{a priori} probabilities of having a 0 or 1 in the raw key. We treat Alice's raw key as the correct one and assume that Bob is correcting his raw key during the error correction step to match Alice's. Thus, it is the \emph{a priori} probabilities of Alice ending up with a 0 or 1 in her raw key that we are interested in. We calculated Alice's \emph{a priori} probabilities of getting a 0 or 1 according to Eq. \ref{eq.APrioriProb}, where $N_{0/1}$ is the number of 0's/1's measured over the course of the experiment which can be computed from the last line in Table \ref{tab.CoinMatrix}.
\begin{equation}\label{eq.APrioriProb}
p_{0/1} = \frac{N_{0/1}}{N_{0} + N_{1}}
\end{equation}
Calculating these, we found $p_{0} = 0.4725$ and $p_{1} = 0.5275$. In order to take care of this imbalance during privacy amplification, we would then have to add a term to Eq. \ref{eq.BitRate} so that it becomes Eq. \ref{eq.BitRateWithAPrioriProb}, where the $h_{2}(p_{0})$ term is the extra information leaked by the unequal \emph{a priori} probabilities of Alice getting a 0 or 1.
\begin{equation}\label{eq.BitRateWithAPrioriProb}
N_{\rm{secure}} = N_{\rm{raw}} (1 - h_{2}(\rm{QBER}) - h_{2}(\rm{p_{0}})) - N_{\rm{leakage}} - N_{\rm{safety}}
\end{equation}
Note that the binary entropy function is symmetric so that it does not matter if we use $h_{2}(p_{0})$ or $h_{2}(p_{1})$ in Eq. \ref{eq.BitRateWithAPrioriProb}. Computing the extra term for our experiment we find that as a first estimate we would have to shrink the final key size by an additional 0.22\% to compensate for the unequal \emph{a priori} probabilities.

\begin{figure}[htbp]
    \centering
    \includegraphics[width=13cm]{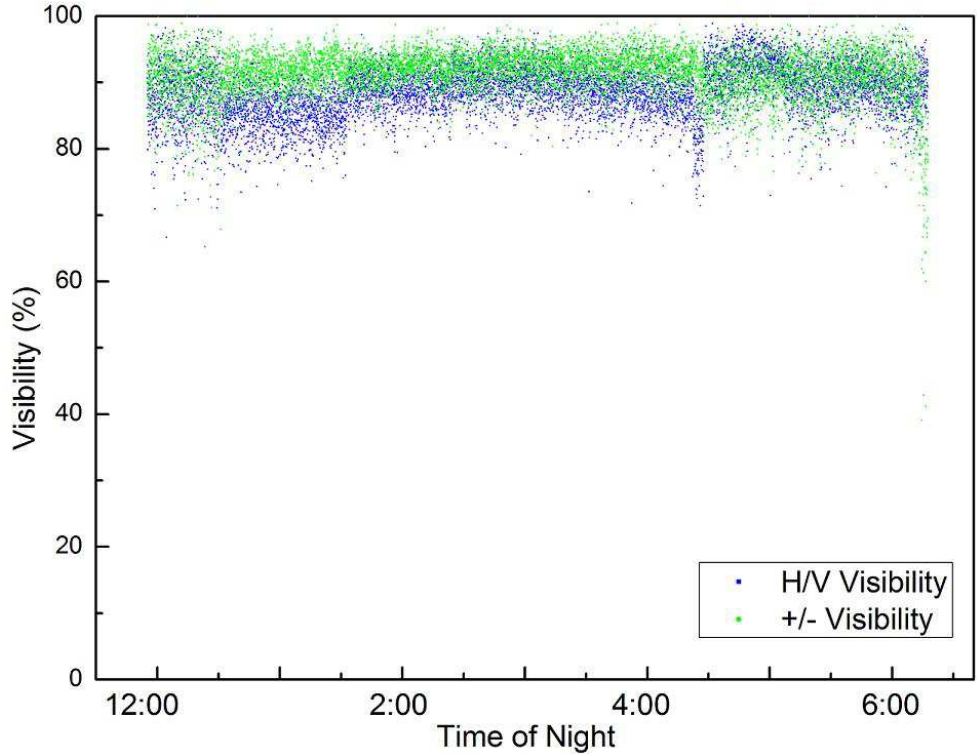}
    \caption{The visibility of the source in the rectilinear (H/V) basis shown in blue and diagonal basis ($+45^{\circ}$/$-45^{\circ}$) shown in green over the course of the experiment. The average visibilities of the two bases recorded during the night were 88.6\% and 91.7\% respectively. Data taken on April 28, 2008. \label{fig.Visibilities}}
\end{figure}

Table \ref{tab.CoinMatrix} also allows one to calculate an average observed visibility of 88.6\% in the H/V basis and 91.7\% in the +/- basis during the experiment. Normally one sees a higher visibility in the H/V basis, not the +/- basis. This is because for the data presented here we had the fiber polarization correction set to map H/V to +/- and vice versa in this experimental run. Nevertheless, this makes no difference to the generation and security of the final key. Fig. \ref{fig.Visibilities} tracks the visibilities in the two bases throughout the experiment.

During the experiment, our modified implementation of the cascade algorithm used average block sizes of 16, 33, 67, 138, and 314 bits for the 5 passes it made over the sifted key data; revealing an average of 174\unit{bits/s}. After error correction, the error corrected key had an average of $1.92 \times 10^{-3}$ residual errors per bit with 8,150 errorless blocks from a total of 9,564 blocks. The somewhat high residual error rate was due to a number of reasons. Simplifications were made in our implementation of the cascade error correction algorithm; namely, rather than going back through all previous passes of cascade the algorithm instead only went back to the first pass. This reduced the effectiveness of our error correction algorithm; however, this was not the dominant source of error since it has been shown that two passes of cascade are usually enough to remove the majority of errors between two bit strings \cite{SY00}. The dominant source of residual error was due to using the error rate estimate, performed by publically revealing 10\% of the sifted key, in order to determine the proper block size for the cascade algorithm. The relatively small sample sizes caused large statistical fluctuations in the error rate estimate leading to a poor choice of the block sizes used in cascade. Improper block sizes in cascade can strongly reduce its effectiveness and were the major source of error in our error correction algorithm. Also, cascade is optimized to work on blocks with errors spread uniformly throughout, in order to accomplish this the sifted key should be randomized before performing cascade on it. Once cascade is properly implemented with efficient block sizes and sifted key randomization the residual error rate can be set to any desired level dependent upon the number of passes performed with cascade.

To test the impact of the block size on the residual error rate we have performed off-line tests by running the saved sifted key data through cascade only this time using either the error rate observed during error correction on the previous block of data or a running average error rate from the last few blocks of data. Already we have seen a significant improvement in the observed residual error rates for the same sifted key data sets. Further work to optimize our implementation of cascade following the analysis of Sugimoto and Yamazaki \cite{SY00} has already begun. Note the fact that the error rate estimation was useless to optimize the block sizes for cascade removes any lingering reasons to even estimate the error rate in the first place and waste 10\% of the key. Theoretical protocols usually describe doing this in order to detect an eavesdropper; however, the error correction algorithm already yields the true error rate which can be used to test the security of the key. So, unless one needs it to optimize the error correction algorithm, there is no reason to do an error rate estimation.

Table \ref{tab.ClassicalComm} shows the average classical communication load in bytes per second during the experiment for coincidence information (Coin Sent and Rec), error rate estimation (ERE Sent and Rec), and error correction (EC Sent and Rec) from Alice's side. In this first implmentation we only made a minimal attempt at optimizing the classical communication load; for example, we sent individual parity bits as a full byte of data. The majority of the communication load was due to sending the timetag information necessary to identify coincident detection events corresponding to the detection of entangled photon pairs. Efforts can be made to ensure that the observer with the lowest detection rates is the one to send the timetag data across; however, beyond that there is not much that can be done to lower the communication load for coincident detection and it should remain the dominant classical communication load in any system that uses timing information to identify entangled photons.

\begin{table}[htbp]
  \centering
  \begin{tabular}{|c|c|c|c|c|c|}
      \hline
      \multicolumn{6}{|c|}{Alice} \\
      \hline
      Coin Sent & Coin Rec & ERE Sent & ERE Rec & EC Sent & EC Rec \\
      \hline
      252,000 & 9,515 & 233 & 8 & 724 & 714 \\
      \hline
  \end{tabular}
  \caption{Classical communication load in bytes/s for the different classical communication tasks needed by the software in order to perform the BBM92 protocol. \label{tab.ClassicalComm}}
\end{table}

Besides the experiment detailed above, experiments were performed with the following combinations of free-space links: a system with completely local detection (used as a baseline), one 435\unit{m} link and local detection, two 435\unit{m} links to adjacent offices in the IQC building, one 1.325\unit{km} link to PI and local detection, a second night's worth of data for the full two link system with one 435\unit{m} and one 1.325\unit{km} free-space link, and the two link experiment detailed above. The data for each experiment including the one detailed above is summarized in Table \ref{tab.OtherExperiments}. From the table it is clear that the wildly varying free-space link to PI
cuts down the raw detection rates significantly, whereas the more stable BFG link shows higher rates. The QBER varied from experiment to experiment depending on how well we were able to compensate for the random polarization rotation induced by the long singlemode fibers which carried the photons to the sender telescopes.

\begin{table}[htbp]
  \centering
  \begin{tabular}{|c|c|c|c|c|c|}
      \hline
       & \multicolumn{4}{|c|}{Key Rates (bits/s)} & \\
      \hline
      Situation & Raw & Sifted & Optimal Final & Actual Final & QBER \\
      \hline \hline
      Local System & 6,025 & 3,052 & 1,894 & - & 2.91\% \\
      \hline
      1 435m BFG link & 2,812 & 1,402 & 656 & - & 4.55\% \\
      \hline
      2 435m BFG links & 1,170 & 582 & 177 & 100 & 6.58\% \\
      \hline
      1 1.35km PI link & 1,398 & 714 & 334 & 244 & 4.58\% \\
      \hline
      435m BFG \& 1.35km PI links \#1 & 857 & 425 & 86 & 34 & 7.98\% \\
      \hline
      435m BFG \& 1.35km PI links \#2 & 565 & 284 & 124 & 85 & 4.92\% \\
      \hline
  \end{tabular}
  \caption{Data for different experimental free-space link setups. \label{tab.OtherExperiments}}
\end{table}

Lastly, a proof-of-principle Bell inequality violation experiment was performed just before the QKD experiment explained in detail above. Over the course of half an hour of data collection we were able to measure an average Bell parameter of 2.51 $\pm$ 0.11, almost 5 standard deviations above the classical limit of 2.

\section{Conclusion}
In conclusion, we have implemented the first real-time two free-space link entangled quantum key distribution system including all error correction and privacy amplification algorithms. The system spans a distance of 1,525\unit{m} with no direct line of sight between Alice and Bob. The source is placed between Alice and Bob with line of sight to each one and takes advantage of the fact that in an entanglement based scheme the source need not be in a trusted location. The system implements the BBM92 protocol and sends pairs of entangled photons over two separate free-space optical links to be detected by Alice and Bob and turned into a secure key. Custom software was written to extract entangled photon detection events using a coincidence detection algorithm rather than relying on timing information from a separate classical channel. Over the course of more than six hours of continuous night time operation the system generated an average raw key rate of 565\unit{bits/s}, sifted key rate of 284\unit{bits/s}, and final secure key rate of 85\unit{bits/s} with an average QBER of 4.92\%.

\section*{Acknowledgments}
Support for this work by NSERC, QuantumWorks, CIFAR, CFI, CIPI, ORF, ORDCF, ERA, and the Bell family fund is gratefully acknowledged. The authors would like to thank: P. Forbes, I. Soellner, N. Ilic, and E. Bocquillon for all of their help with late night data collection and system testing; B. Schmidt for his initial programming of the error correction and privacy amplification routines; M. Peloso for his design of the detector box and of the initial free-space optics; R. Horn, D. Smith, M. Laforest, and R. Kaltenbaek for their construction help, optics advice, and testing help; R. Irwin, T. Gerhardt, J. Thompson, and P. McGrath; N. L\"utkenhaus, H.K. Lo, and K. Resch for their comments on early drafts of this paper and their help, encouragement, and enlightening discussions about the project; the anonymous referees for their many comments, which were very useful in improving the quality of this paper; and Herb Epp and the Environment and Parks Department of the City of Waterloo for removing a particularly annoying arboreal eavesdropper that was performing a denial-of-service attack on our system.

\end{document}